\documentclass[lettersize,journal]{IEEEtran}
\usepackage{amsmath,amsfonts}
\usepackage{algorithmic}
\usepackage{algorithm}
\usepackage{array}
\usepackage[caption=false,font=normalsize,labelfont=sf,textfont=sf]{subfig}
\usepackage{textcomp}
\usepackage{stfloats}
\usepackage{url}
\usepackage{multirow}
\usepackage{verbatim}
\usepackage{graphicx}
\usepackage{makecell}
\usepackage{tabularx}
\usepackage{booktabs}
\usepackage{makecell}
\usepackage{cite}
\usepackage{enumitem}
\usepackage{pgfplots}
\pgfplotsset{compat=1.18}
\begin{document}

\title{Timbre-Aware LLM-based Direct Speech-to-Speech Translation Extendable to Multiple Language Pairs}
\author{
Lalaram~Arya,
Mrinmoy~Bhattacharjee,
Adarsh~C.~R.,
and~S.~R.~Mahadeva~Prasanna%
\thanks{Corresponding author: Mrinmoy Bhattacharjee (e-mail: mrinmoy.bhattacharjee@iitjammu.ac.in).}
\thanks{
Lalaram Arya, Adarsh C. R., and S. R. Mahadeva Prasanna are with the
Indian Institute of Technology Dharwad, Dharwad, India
(e-mail: 202021004@iitdh.ac.in; adarshcr@iitdh.ac.in; prasanna@iitdh.ac.in).
}
\thanks{
Mrinmoy Bhattacharjee is with the
Indian Institute of Technology Jammu, Jammu, India
(e-mail: mrinmoy.bhattacharjee@iitjammu.ac.in).
}
}




\maketitle

\begin{abstract}
Direct Speech-to-Speech Translation (S$2$ST) has gained increasing attention for its ability to translate speech from one language to another, while reducing error propagation and latency inherent in traditional cascaded pipelines. However, existing direct S$2$ST systems continue to face notable challenges, including instability in semantic–acoustic alignment when parallel speech data is scarce, difficulty in preserving speaker identity, and limited multilingual scalability. In this work, we introduce DS$2$ST-LM, a scalable, single-stage direct S$2$ST framework leveraging a multilingual Large Language Model (LLM). The architecture integrates a Whisper speech encoder, a learnable projection module, a Qwen$2$--$0.5$B LLM, and a timbre-controlled vocoder. We construct GigaS$2$S-$1000$, a $1000$-hour bilingual corpus by extending the GigaST dataset with high-fidelity synthetic target speech, and show that this synthetic data alleviates data scarcity to some extent. We investigate two semantic token generation strategies: speech-derived $S^3$ tokens and text-derived tokens generated by a pre-trained LLM, and analyze their impact on training stability and semantic consistency. We further evaluate three projection architectures (Linear, Conv$1$D–Linear, and Q-Former) and observe that while higher-capacity projectors converge faster, the simple Linear projector achieves higher performance. Extensive experiments demonstrate that DS$2$ST-LM outperforms traditional cascaded and ST (Qwen-Audio) $+$ TTS baselines across both lexical (BLEU, METEOR) and semantic (BLEURT, COMET) metrics, while extending to multiple language pairs, including French, Spanish, German, Hindi, Bengali, and Urdu. Furthermore, we incorporate timbre-aware speech synthesis to preserve speaker information, enabling DS$2$ST-LM to surpass prior direct S$2$ST systems in both speaker similarity and perceptual naturalness.
\end{abstract}

\section{Introduction}
The rapid growth of global communication over the past few decades has increased interaction among people speaking diverse languages. According to recent estimates~\footnote{https://www.ethnologue.com/}, approximately $7{,}159$ languages are currently in use worldwide, highlighting the scale of linguistic diversity and the associated communication challenges. Speech-to-Speech Translation (S$2$ST)~\cite{nakamura2006atr} has therefore emerged as a promising technology for bridging these language barriers. Conventional S$2$ST systems typically adopt a cascaded architecture composed of Automatic Speech Recognition (ASR)~\cite{alharbi2021asr}, Machine Translation (MT)~\cite{tan2020nmt}, and Text-to-Speech (TTS)~\cite{barakat2024expressive} modules. While such systems achieve acceptable performance in high-resource settings, they suffer from error propagation across modules, loss of prosodic and paralinguistic information~\cite{le2024transvip}, increased latency due to multi-stage processing, and difficulties in translating languages that are predominantly available in spoken form and do not have a written script~\cite {Etchego2022}~\cite{zhang2020uwspeech}.

To mitigate some of these limitations, researchers have explored end-to-end Speech-to-Text (ST) models that directly map source-language speech to target-language text, followed by TTS-based speech synthesis~\cite{weiss2017seq2seq}. 
Although these pipelines reduce error propagation compared to cascaded architectures, they still require the independent training of the ST and TTS modules. In addition, integrating these modules and the availability of target-language text is necessary~\cite{dabre2024nict}. Subsequent efforts have attempted to pretrain ST and TTS independently and connect them through an additional fine-tuning stage~\cite{fang2023daspeech}. While this approach enables direct inference from source speech to target speech, it remains text-dependent and computationally expensive~\cite{inaguma2023unity}.

More recent research has therefore focused on fully end-to-end S$2$ST frameworks that jointly learn the mapping from source-language speech to target-language speech within a single model~\cite{jia2019direct}. These approaches initially relied on attention-based sequence-to-sequence architectures, with later work incorporating pretrained speech encoders and neural vocoders to improve translation quality and speech naturalness~\cite{popuri2022enhanced_s2st}. Despite these advances, current end-to-end S$2$ST models often lag behind strong cascaded baselines in translation quality, even though in reduced error propagation~\cite{lee2022direct}.

In parallel with these developments, large language models (LLMs) have demonstrated strong capabilities across a wide range of speech and language processing tasks~\cite{cui2025slm}. Emerging LLMs for speech further extend these capabilities to spoken question answering, ASR, TTS, and ST~\cite{xia2024cotst}~\cite{zhang2023vallex}. 
However, the potential of LLMs for direct S$2$ST remains largely underexplored~\cite{borsos2023audiopalm}. At the same time, progress in spoken dialogue systems shows that LLMs can support speech-based interaction when equipped with audio adapters and neural vocoders~\cite{chen2025slamomni}. These advances highlight promising opportunities for building scalable LLM-driven direct S$2$ST systems, although most prior work remains limited to preliminary explorations using synthetic data~\cite{xu2025slamtr}.

Motivated by these recent advances, we develop a Direct Speech-to-Speech Translation System with LLM (DS$2$ST-LM), a scalable, single-stage direct S$2$ST framework that leverages the language understanding capabilities of LLMs. While direct S$2$ST systems require parallel speech pairs, such data remain scarce for many languages. To address this challenge, we construct GigaS$2$S-$1000$, a large-scale corpus derived from GigaST~\cite{tang2023gigast} by synthesizing high-quality Chinese speech using XTTS-v$2$~\cite{casanova2024xtts}. The resulting corpus contains $1000$ hours of multi-speaker English speech aligned with high-quality machine-translated Chinese text and natural-sounding single-speaker Chinese speech, enabling large-scale direct S$2$ST training.

DS$2$ST-LM integrates a Whisper encoder~\cite{radford2023whisper}, a Qwen$2$--$0.5$B LLM~\cite{yang2024qwen2}, and a CosyVoice neural vocoder~\cite{du2024cosyvoice}. 
Since LLMs operate in a text embedding space, a projector module is used to interface the encoder's speech embedding space with that of the LLM. We investigate three projection designs: Linear~\cite {chen2025slamomni}, Conv$1$D–Linear~\cite{song2025phoneme}, and Q-Former~\cite{li2023blip2}, and analyze their impact on training and translation quality. 
The vocoder synthesizes target speech from semantic tokens. These semantic tokens are discrete representations that primarily capture linguistic and contextual information, unlike acoustic tokens, and can be generated from either target speech or text.
Accordingly, we explore two training regimes: one that uses semantic tokens extracted directly from target speech and another that uses tokens generated from target text via a text-to-token LLM. We evaluate both training regimes and compare their effects on translation performance. 

We further demonstrate the extensibility of DS$2$ST-LM across multiple language pairs translated into English (en), including French (fr), Spanish (es), and German (de), using the CVSS corpus~\cite{jia2022cvss}. We also evaluate Indic language pairs, including Hindi (hi), Bengali (ben), and Urdu (urd), using the Bhasaanuvaad dataset~\cite{jain2025bhasaanuvaad}. Finally, we incorporate a timbre-control mechanism inspired by recent TTS systems~\cite{du2024cosyvoice,casanova2024xtts}, which enables DS$2$ST-LM to synthesize target speech in a specified speaker’s voice from a short reference audio prompt.

The main contributions of this work are summarized as follows:
\begin{itemize}
    \item We propose DS$2$ST-LM, a single-stage, LLM-based direct S$2$ST framework, and demonstrate its effectiveness across multiple language pairs.
    \item We construct and publicly release the GigaS$2$S-$1000$ dataset, enabling large-scale research on direct S$2$ST.
    \item We systematically evaluate three projection architectures: Linear, Conv$1$D–Linear, and Q-Former, and analyze their effects on convergence stability and translation quality.
    \item We investigate semantic token generation from target speech versus target text and analyze their impact on direct S$2$ST translation performance.
    \item We integrate timbre-aware speech synthesis into direct S$2$ST by conditioning on semantic tokens and a reference speaker prompt to synthesize speaker-specific target speech.
    \item We release all training recipes, evaluation pipelines, and model checkpoints to support reproducibility and future research.
\end{itemize}

\section{Background}
This section provides an overview of the evolution of S$2$ST systems, tracing the progression from early rule-based models to current large-scale, end-to-end neural architectures. Broadly, this evolution can be categorized into three phases: (i) cascaded S$2$ST systems, including ASR~\cite{alharbi2021asr}–MT~\cite{tan2020nmt}–TTS~\cite{barakat2024expressive} pipelines and ST~\cite{weiss2017seq2seq}-to-TTS cascades; (ii) direct S$2$ST~\cite{jia2019direct} models based on sequence-to-sequence architectures; and (iii) recent approaches that leverage LLMs for speech understanding and generation~\cite{borsos2023audiopalm}. We begin by discussing cascaded S$2$ST systems, which laid the groundwork for subsequent direct and LLM-based methods.

\subsection {Cascaded speech-to-speech translation}
Cascaded S$2$ST systems traditionally integrate three sequential modules: ASR, MT, and TTS, to translate source-language speech into target-language speech. Historically, this paradigm evolved from early rule-based prototypes to corpus-driven and neural architectures~\cite{dhawan2022S2ST}. Early demonstrations by NEC Corporation and later collaborations among Advanced Telecommunications Research (ATR), CMU, and Siemens established the feasibility of real-time S$2$ST~\cite{seligman2017taus,morimoto1993asura,kurematsu1993overview}. The JANUS system formalized a modular pipeline for speech understanding and translation~\cite{waibel1991janus}, which was later extended to multilingual and multidomain settings through statistical MT~\cite{lavie1997janus}. The Verbmobil project further advanced conversational S$2$ST by integrating syntactic parsing, semantic transfer, and dialogue management~\cite{wahlster2000verbmobil}.

Building on these foundations, more comprehensive cascaded systems emerged. The ATR multilingual S$2$ST pipeline demonstrated a complete system for travel-related conversations~\cite{nakamura2006atr}. Subsequent initiatives such as TC-STAR~\cite{lazzari2006tcstar} and GALE~\cite{liu2010gale} expanded cascaded S$2$ST to open-domain and broadcast speech by leveraging large-scale data and enhanced ASR~\cite{alharbi2021asr}, MT~\cite{tan2020nmt}, and TTS~\cite{barakat2024expressive} modules. With the emergence of deep learning, cascaded systems incorporated neural acoustic encoders~\cite{hannun2014deepspeech}, Transformer-based MT~\cite{vaswani2017attention}, and neural vocoders~\cite{oord2016wavenet}. Industrial-scale deployments, including Google’s Neural Speech Translator~\cite{guo2022hwtsc} and Huawei’s HW-TSC series~\cite{wang2023hwtsc}, further improved prosody preservation and speaker timbre transfer, advancing natural and low-latency multilingual speech translation.

Although cascaded systems remain widely used due to their modularity, they suffer from inherent limitations. Error propagation across ASR, MT, and TTS stages degrades translation accuracy, while reliance on intermediate transcriptions limits scalability. These constraints have motivated more integrated approaches. End-to-end speech-to-text (ST) models demonstrated the feasibility of directly predicting target-language text from source speech using attention-based encoder–decoder architectures~\cite{berard2016listen}. Conformer-based encoders later improved robustness by better modeling local speech features~\cite{gulati2020conformer}. More recently, LLM-based approaches have been introduced into ST pipelines. CoT-ST~\cite{xia2024cotst} incorporates multimodal chain-of-thought reasoning to improve contextual interpretation, while LLaST~\cite{chen2024llast} enhances multilingual ST through ASR-augmented training and parameter-efficient adaptation.
\subsection {Direct Speech-to-Speech Translation}
Direct S$2$ST aims to generate target speech directly from source speech without relying on intermediate text, thereby reducing latency and avoiding error accumulation typical of cascaded pipelines~\cite{jia2019direct}. Early work demonstrated that attention-based encoder–decoder architectures could map source speech spectrograms to target acoustic features, establishing the feasibility of fully end-to-end speech translation~\cite{jia2022translatotron2}. Subsequent studies strengthened these models through self-supervised pretraining and by disentangling linguistic and speaker representations to improve semantic mapping and speaker preservation~\cite{lee2022textless}. Building on these foundations, discrete and textless modeling approaches replaced continuous spectrograms with sequences of discrete speech units derived from self-supervised encoders such as HuBERT~\cite{lee2022direct}. These unit-based representations improved translation robustness and cross-lingual scalability, enabling S$2$ST without intermediate text~\cite{lee2023textlesslow}. Further extensions introduced many-to-many multilingual direct S$2$ST via unit-to-unit modeling, allowing translation across multiple language pairs entirely in the discrete acoustic domain~\cite{kim2024textless}. In parallel, direct S$2$ST has been explored for unwritten and low-resource languages, where intermediate unit representations and vector-quantized variational autoencoder (VQ-VAE) models are used to reconstruct intelligible target speech without text supervision~\cite{zhang2020uwspeech,chen2023unwritten}.

\subsection {Direct Speech-to-Speech translation using LLM}
Recent progress in direct S$2$ST has increasingly shifted toward frameworks built around large language models (LLMs). Early studies demonstrated that decoder-only LLMs can be adapted for speech understanding and generation, allowing models to retain strong semantic reasoning capabilities acquired through large-scale text pretraining~\cite{rubenstein2023audiopalm}. VALL-E X~\cite{zhang2023vallex} explored a cross-lingual neural codec language model that preserves speaker identity and generates natural prosody across languages via joint speech–text conditioning in a zero-shot prompt setup. Building on these ideas, subsequent work proposed unified multimodal pipelines that integrate pretrained speech encoders with LLM-based decoders through a projection module, followed by neural vocoders, leading to improved semantic alignment and contextual coherence in conversational speech tasks~\cite{chen2025slamomni,fang2025llamaomni2}. More recent efforts have further explored direct S$2$ST by leveraging synthetic data and jointly training speech encoders, modality projectors, LLM decoders, and vocoders to enhance translation accuracy~\cite{xu2025slamtr}. Despite this progress, several aspects remain underexplored in LLM-based S$2$ST, including the impact of natural speech data on translation quality, the role of different projection architectures, speaker prosody transfer, and scalability to a broader range of languages.
\section{Proposed Work}
In this section, we present an LLM-based framework for direct S$2$ST that jointly models speech understanding, cross-lingual semantic translation, and speech generation within a unified architecture. We first describe GigaS$2$ST-$1000$, a large-scale dataset constructed for direct S$2$ST training, followed by a speech tokenization strategy that represents speech as semantically aligned discrete tokens. Building on these representations, we introduce a direct S$2$ST framework comprising a speech encoder, a cross-modal projection module, and a multilingual LLM decoder that jointly predict semantic speech tokens and text in the target language. Finally, a timbre-controllable neural vocoder conditioned on speaker embeddings synthesizes high-quality target speech while preserving speaker characteristics.

\subsection{Development of GigaS$2$S-$1000$}\label{gigas2s}
The GigaS$2$S-$1000$\footnote{https://huggingface.co/datasets/Lalaramarya/GigaS2S-1000} dataset is constructed using the GigaSpeech~\cite{chen2021gigaspeech} and GigaST~\cite{tang2023gigast} corpora. GigaSpeech, originally developed for ASR, contains large-scale English speech collected from diverse online sources, including audiobooks, podcasts, and YouTube, spanning $24$ broad thematic categories such as education, science and technology, business, sports, and entertainment. This diversity introduces substantial variation in speaking style, topic distribution, and acoustic conditions, making the corpus suitable for training robust speech translation models. Each audio recording is paired with carefully verified utterance-level transcriptions obtained by segmenting long-form recordings.

GigaST~\cite{tang2023gigast} extends GigaSpeech for English-to-Chinese speech translation by translating the English transcripts using a Transformer-based context-aware MT system. The MT model is trained on large multilingual text corpora, including WMT2021, CCMatrix, CCAlign, and OpenSubtitles from OPUS~\cite{tiedemann2012opus}. To further improve the quality of GigaS$2$S-$1000$, English–Chinese sentence pairs are filtered using SONAR embedding similarity, retaining only samples with cosine similarity above $0.9$. This process produces a highly semantically aligned $1000$-hour subset of GigaST, which is used to synthesize corresponding Chinese speech, yielding a cleaner, more reliable dataset for large-scale direct S$2$ST training.

\subsubsection{Audio generation using TTS}
Although the GigaST~\cite{tang2023gigast} corpus provides English source speech, English transcriptions, and corresponding translated Chinese text, it does not include aligned Chinese speech. To enable its use for S$2$ST research, we extend GigaST by synthesizing the Chinese speech. 

To generate the speech data, we employ XTTS-v$2$~\cite{casanova2024xtts}, a state-of-the-art multilingual TTS system that supports zero-shot voice adaptation using a $6$-second speech prompt. XTTS-v$2$ uses a GPT-$2$–based encoder to generate discrete speech representations, which are converted into high-fidelity waveforms via a HiFi-GAN vocoder~\cite{kong2020hifigan}. Trained on large-scale multilingual datasets, the model supports $17$ languages and enables cross-lingual voice cloning, style transfer, and emotional adaptation, producing high-quality speech at a $24$-kHz sampling rate. These properties allow effective generalization to unseen speakers and languages, making XTTS-v$2$ well-suited for synthesizing speech for S$2$ST training.

Using XTTS-v$2$, we synthesize Chinese speech and construct the GigaS$2$ST-$1000$ corpus, which contains approximately $833{,}000$ bilingual sentence pairs and around $1{,}000$ hours of aligned speech. The original English speech, collected from diverse web sources, includes recordings from multiple male and female speakers, whereas the Chinese speech is synthesized using a single male speaker's voice. Both English and Chinese speech are provided at a $16$-kHz sampling rate with mono-channel audio. The dataset is publicly released on Hugging Face\footnotemark[\value{footnote}] and includes English speech with transcriptions, translated Chinese text, and natural-sounding synthetic Chinese speech.

\subsection{Speech Tokenization}\label{token}
Speech tokenization is a primary step in current speech and language processing, converting continuous speech signals into sequences of discrete tokens. These tokens can be categorized into acoustic and semantic tokens. Acoustic tokens, generated using neural audio codecs with vector quantization, preserve timbre and high-fidelity representations suitable for tasks such as speech synthesis and compression~\cite{zeghidour2021soundstream}, whereas semantic tokens are derived from self-supervised speech models that capture linguistic and contextual information~\cite{hsu2021hubert}.

However, self-supervised tokenization methods may exhibit weak alignment with semantics, leading to incomplete encoding of linguistic meaning~\cite{hsu2021hubert}. To address this limitation, a supervised semantic tokenization approach is developed to extract tokens from multilingual ASR encoders coupled with vector quantization~\cite{du2024cosyvoice}, enabling a more faithful mapping between speech features and textual semantics and producing semantically aligned tokens better suited for S$2$ST tasks.
\begin{figure}[!t]
\centering
\includegraphics[width=0.45\textwidth]{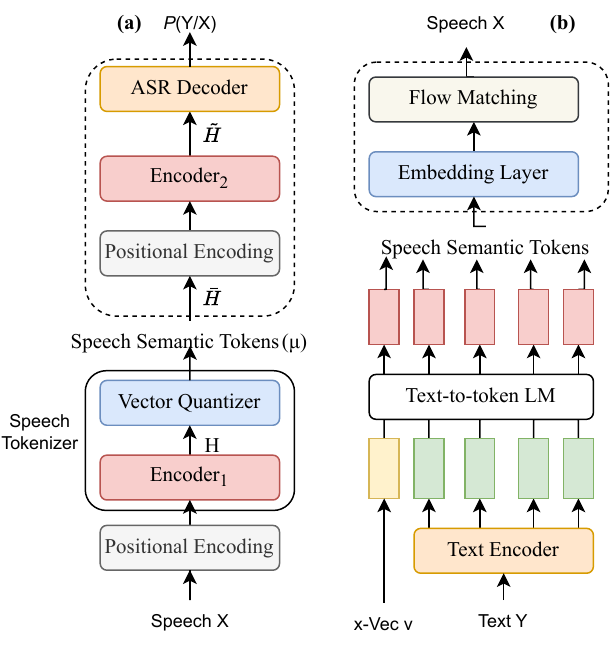}
\caption{Overview of semantic token generation using the $\mathcal{S}^{3}$ tokenizer from speech (left) and a text-to-token LLM-based approach from text (right). Dashed modules denote components used only during training.}
\label{s3}
\end{figure}

\subsubsection{Supervised Semantic Token Generation from Speech}
To generate supervised semantic tokens from speech, we adopt the framework proposed in~\cite{du2024cosyvoice}, in which a multilingual pre-trained SenseVoice ASR encoder is split into two parts with a vector quantization (VQ) module inserted between them. The first six layers are denoted as $\mathrm{Encoder}_{1}$, and the remaining layers as $\mathrm{Encoder}_{2}$, as illustrated in Fig.~\ref{s3}(a). Given an input speech signal, the mel-spectrogram feature sequence $\mathbf{X}$ is passed through positional encoding and $\mathrm{Encoder}_{1}$ to produce a context-aware representation $\mathbf{H}$:
\begin{equation}
\mathbf{H} = \operatorname{Encoder}_{1}(\operatorname{PosEnc}(\mathbf{X}))
\label{eq1}
\end{equation}
where the positional encoding operation is represented by $\operatorname{PosEnc}(\cdot)$. Each frame-level representation $\mathbf{h}_i \in \mathbf{H}$, for $i = 1, \ldots, T$, where $T$ is the number of speech frames, is then quantized using a vector quantizer to obtain a discrete token $\mu_i$, corresponding to the index of the nearest embedding vector in the codebook $\mathcal{C} = \{\mathbf{c}_1, \ldots, \mathbf{c}_{|\mathcal{C}|}\}$:
\begin{equation}
\mu_i = \mathrm{VQ}(\mathbf{h}_i) 
= \arg \min_{\mathbf{c}_n \in \mathcal{C}} \|\mathbf{h}_i - \mathbf{c}_n\|_2
\label{eq2}
\end{equation}
where $\|\cdot\|_2$ denotes the $\ell_2$ norm. The corresponding codebook embeddings $\bar{\mathbf{H}} = [\mathbf{e}_{\mu_1}, \mathbf{e}_{\mu_2}, \ldots, \mathbf{e}_{\mu_T}]$ are subsequently processed by $\mathrm{Encoder}_{2}$ to generate the latent representation $\tilde{\mathbf{H}}$:
\begin{equation}
\tilde{\mathbf{H}} = \operatorname{Encoder}_{2}(\operatorname{PosEnc}(\bar{\mathbf{H}}))
\label{eq3}
\end{equation}
Finally, $\tilde{\mathbf{H}}$ is fed into a Transformer-based ASR decoder to predict the posterior probabilities of the target text sequence:
\begin{equation}
P(\mathbf{Y} \mid \mathbf{X})  = \operatorname{ASRDecoder}(\mathbf{Y}^{Z-1},\tilde{\mathbf{H}})
\label{eq4}
\end{equation}
where $\mathbf{Y}^{Z-1}$ denotes the left-shifted target text sequence used in supervised training.

\subsubsection{Semantic Token Generation from Text}
Direct S$2$ST is an emerging research area, and publicly available S$2$ST datasets remain limited. In contrast, a large amount of ST data is available and can be effectively leveraged to train direct S$2$ST systems. However, ST datasets typically do not include target-language speech. In such cases, semantic tokens can be generated from the target-language text using the text-to-token LLM provided in CosyVoice-300M-SFT~\cite{du2024cosyvoice}. As illustrated in Fig.~\ref{s3}(b), the LLM generates semantic tokens autoregressively conditioned on the textual input.

During training, a teacher-forcing strategy is employed, conditioning the model on the target text, a speaker embedding extracted from a short reference prompt, and the ground-truth semantic token sequence. This design enables the LLM to learn a stable text-to-semantic mapping. It allows the direct S$2$ST system to exploit large-scale ST corpora even when paired target-language speech is unavailable.

\subsection{LLM-Based Speech-to-Speech Translation Framework} 
\begin{figure}[!t]
\centering
\includegraphics[width=0.48\textwidth]{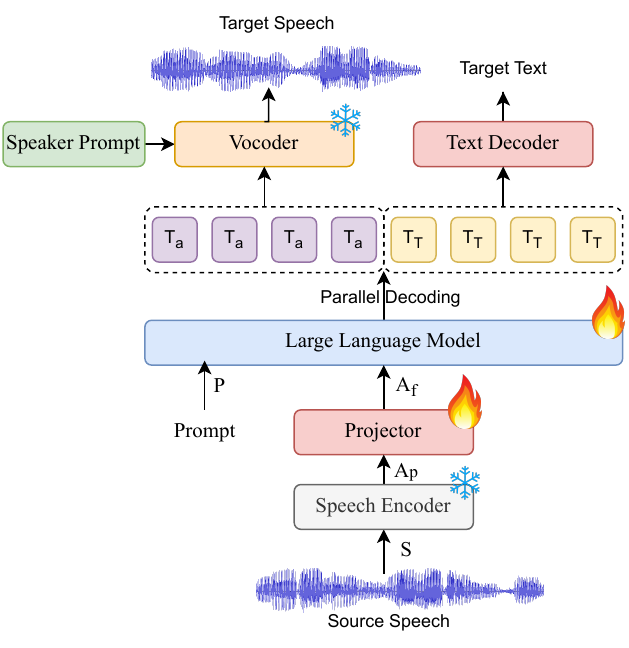}
\caption{Overview of the DS$2$ST-LM framework for the direct S$2$ST task.}
\label{DS2ST-LM}
\end{figure}

DS$2$ST-LM is an end-to-end, single-stage, trainable direct S$2$ST framework inspired by the SLAM-Omni~\cite{chen2025slamomni} architecture, initially proposed for conversational tasks. As illustrated in Fig.~\ref{DS2ST-LM}, the system comprises a speech encoder, a projection module, an LLM decoder, and a neural vocoder. The speech encoder transforms input audio into a sequence of representations capturing high-level acoustic information, which are then mapped by the projection module into the LLM’s embedding space. Operating on these projected embeddings, the LLM autoregressively predicts semantic speech tokens and auxiliary target-language text. The predicted semantic tokens, together with a reference speaker prompt, are subsequently passed to the vocoder to synthesize high-quality target speech while preserving the desired speaker timbre.

\subsubsection{Speech Encoder}
The speech encoder serves as the front end of the system, converting raw audio into high-level acoustic representations compatible with downstream components. We adopt the Whisper encoder from OpenAI, pre-trained on approximately $680{,}000$ hours of multilingual and multi-accent speech, which enables robust modeling of phonetic and linguistic cues and makes it well suited for extendable S$2$ST. In our framework, only the Whisper encoder part is used.
Given an input speech sequence $S = [s_1, s_2, \ldots, s_T]$, the encoder produces a sequence of frame-level feature vectors $\mathbf{A}_f$:
 \begin{equation}
 \mathbf{A}_f = \mathrm{Encoder}_{\mathrm{S2ST}}(S)
 \end{equation}
 These features typically have higher dimensionality than the embedding space expected by the LLM; therefore, a projection module is required to map them into LLM-compatible representations.
\subsubsection{Projection module}
The projection module ensures dimensional consistency and cross-modal alignment between the speech encoder and the LLM, enabling the LLM to process speech-derived representations without disrupting its pre-trained text representations. Mathematically, this can be written as:
\begin{equation}
\mathbf{A}_p = \mathrm{Proj}(\mathbf{A}_f)
\end{equation}
where $\mathbf{A}_f$ denotes the acoustic features generated by the encoder and $\mathbf{A}_p$ represents the projected embeddings. In this work, we explore three projection strategies with varying modeling capacities: linear, hybrid Conv$1$D–Linear, and transformer-based designs such as Q-Former.

{(i) \textbf{Linear}:} The linear projection module~\cite{chen2025slamomni} performs temporal downsampling by grouping every $k$ consecutive encoder frames and applying a lightweight two-layer MLP with ReLU activation. This reduces sequence length while preserving short-range temporal structure and produces LLM-compatible embeddings with minimal computational overhead. Mathematically, this can be written as:
\begin{equation}
\mathbf{A}_p = \mathrm{MLP}\!\left(\mathrm{Group}_k(\mathbf{A}_f)\right)
\end{equation}
where $\mathrm{Group}_k(\cdot)$ concatenates every $k$ consecutive frame-level representations along the temporal dimension.

{(ii) \textbf{Conv$1$D–Linear}:} To enable more flexible temporal aggregation, the Conv$1$D–Linear projector~\cite{song2025phoneme} replaces fixed frame grouping with a one-dimensional convolution using kernel size and stride $k$. The convolution captures local temporal patterns and performs learned downsampling, while the subsequent MLP maps the aggregated features to the target embedding dimension, providing improved temporal consistency and adaptivity over linear projection. Mathematically, this can be written as:
\begin{equation}
\mathbf{A}_p = \mathrm{MLP}\!\left(\mathrm{Conv1D}_k(\mathbf{A}_f)\right)
\end{equation}
where $\mathrm{Conv1D}_k(\cdot)$ denotes a one-dimensional convolution with kernel size and stride $k$ along the temporal dimension.

{(iii) \textbf{Q-Former}:} The Q-Former projector~\cite{li2023blip2}, inspired by BLIP-2, uses transformer-based cross-attention to achieve richer alignment between speech and language representations. A set of learnable query embeddings attends to the speech encoder hidden states to extract semantically informative features, which are then projected into the LLM embedding space, yielding content-aware and globally aligned representations. Mathematically, this can be written as:
\begin{equation}
\mathbf{A}_p = \mathrm{Proj}\!\left(\mathrm{Q\text{-}Former}(\mathbf{Q}, \mathbf{A}_f)\right)
\end{equation}
where $\mathbf{Q} \in \mathbb{R}^{N_q \times d_q}$ denotes a set of $N_q$ learnable query embeddings with hidden dimension $d_q$.

\subsubsection{LLM-based Decoder}
\label{group-modelling}
The projected acoustic representations $\mathbf{A}_p$ are concatenated with a textual prompt $\mathbf{P}$ and fed into the LLM to perform cross-lingual mapping and generate target semantic tokens. The multimodal input is defined as
\begin{equation}
\mathbf{Z} = \mathbf{P} \oplus \mathbf{A}_p
\end{equation}
where $\mathbf{Z}$ denotes the input processed by the LLM. In our system, we employ the Qwen$2$ multilingual LLM~\cite{yang2024qwen2} and use the Qwen$2$–$0.5$B variant due to its lower computational cost.

Given the concatenated input $\mathbf{Z}$, the LLM predicts both semantic speech tokens and text tokens in parallel using autoregressive decoding, with semantic tokens subsequently decoded by the neural vocoder. To enable this dual-modality prediction, the original text vocabulary $V_t$ is augmented with a semantic token codebook $V_a$. Since these semantic tokens are not present in the pretrained LLM, the corresponding embeddings for $V_a$ are randomly initialized, while the pretrained text embeddings associated with $V_t$ remain unchanged to preserve linguistic competence. A key challenge arises from the mismatch between speech token frequency (typically $50$ Hz) and text token frequency (approximately $3$ Hz)~\cite{chen2025slamomni}, which leads to higher latency in speech generation. To address this issue, we adopt the semantic group modeling strategy from SLAM-Omni~\cite{chen2025slamomni}, where audio logits $L_a$ are projected into groups of size $G$, yielding group-level logits $L_g \in \mathbb{R}^{G \times |V_a|}$.

During training, the semantic token sequence $\boldsymbol{\mu}^{T} = [\mu_0, \mu_1, \ldots, \mu_{T-1}]$ is partitioned into grouped sequences $\mathbf{G}^{T'} = [g_0, g_1, \ldots, g_{T'-1}]$, where each group $g_i$ contains $G$ consecutive tokens and is defined as
\begin{equation}
g_{i}=\left[\mu_{i \cdot G}, \mu_{i \cdot G+1}, \ldots, \mu_{(i+1) \cdot G-1}\right] \quad T' = \left\lfloor \frac{T}{G} \right\rfloor
\end{equation}
Conditioned on the prompt embeddings $\mathbf{P}$, projected acoustic features $\mathbf{A}^{P}$, and target text sequence $\mathbf{T}^{L} = [t_0, t_1, \ldots, t_{L-1}]$, the model is trained using a weighted sum of audio-token and text-token cross-entropy losses:
\begin{equation}
\mathcal{L}=\lambda_{\text{audio}} \mathcal{L}_{\text {audio }} + \lambda_{\text {text }} \mathcal{L}_{\text {text }}
\end{equation}
where $\lambda_{\text{audio}}$ and $\lambda_{\text{text}}$ are scalar coefficients controlling the relative contributions of the two objectives and are set to $1$ to ensure equal weighting. The audio token prediction loss $\mathcal{L}_{\text{audio}}$ is defined as
\begin{equation}
\mathcal{L}_{\text {audio }}=-\frac{1}{T^{\prime} G} \sum_{i=0}^{T^{\prime}-1} \sum_{j=0}^{G-1} \log p\left(\mu_{i \cdot G+j} \mid \mathbf{P}, \mathbf{A}^{P}, \mathbf{G}_{<i}^{T'}, \mathbf{T}_{<}^{L}\right)
\end{equation}
and text token prediction loss $\mathcal{L}_{\mathrm{text}}$ is is given by
\begin{equation}
\mathcal{L}_{\mathrm{text}}=-\frac{1}{L} \sum_{i=0}^{L-1} \log p\left(t_{i} \mid \mathbf{P}, \mathbf{A}^{P}, \mathbf{G}_{<}^{T'}, \mathbf{T}_{<i}^{L}\right)
\end{equation}

\subsubsection{Timbre-controllable Vocoder}
Direct S$2$ST systems often rely on discrete units extracted from self-supervised speech models and converted into waveforms using a neural vocoder. These units entangle linguistic content with acoustic and prosodic attributes such as pitch, rhythm, and energy. As a result, utterances with identical semantics may be represented by different unit sequences, which can degrade translation accuracy and speaker consistency. In the proposed DS$2$ST-LM framework, the LLM predicts semantic tokens that explicitly exclude acoustic characteristics. However, since semantic tokens alone cannot reconstruct speaker timbre, an explicit timbre-conditioning mechanism is incorporated during synthesis. Following recent TTS systems~\cite{du2024cosyvoice}, we employ a conditional flow-matching model to generate mel-spectrograms conditioned on the LLM-predicted semantic tokens and a speaker embedding extracted from a short reference speaker prompt. These mel-spectrograms are then converted into waveforms using a HiFi-GAN vocoder~\cite{kong2020hifigan}, enabling high-fidelity speech synthesis with controllable speaker timbre.

\section{Experiments and Results}
In this section, we describe the datasets and baseline systems used in our study, followed by the training setup, evaluation metrics, and a comprehensive evaluation of the experimental outcomes for the proposed DS$2$ST-LM framework.

\subsection{Dataset}
We utilize four datasets to train and evaluate the  DS$2$ST-LM system: GigaS$2$ST-$1000$, Seamless-Align~\cite{barrault2023seamlessm4t}, CVSS~\cite{jia2022cvss}, and the Bhasaanubad~\cite{jain2025bhasaanuvaad}. As detailed in Section~\ref{gigas2s}, the GigaS$2$ST-$1000$ dataset developed in this work comprises $1000$ hours of high-quality, semantically aligned zh--en speech data, suitable for direct S$2$ST training. Seamless-Align~\cite{barrault2023seamlessm4t}, released by Meta AI for multilingual S$2$ST research, provides large-scale aligned natural speech recordings across $30$ languages; in our experiments, we use the zh--en subsets, totaling approximately $1000$ hours of speech data. Since Seamless-Align does not include parallel transcriptions for both languages, we generate English transcriptions using the Whisper-Large ASR model~\cite{radford2023whisper} for evaluation.

We further incorporate the CVSS dataset~\cite{jia2022cvss}, a multilingual S$2$ST corpus derived from the CommonVoice and CoVoST2 datasets. CVSS includes two variants: CVSS-C, which provides single-speaker synthesized target speech, and CVSS-T, which contains voice-cloned target speech matching the source speaker. In this work, we use CVSS-C and select the fr, es, and de subsets, together comprising approximately $384$ hours of speech paired with English translations. To extend our study to Indic languages, we utilize the Bhasaanubad dataset~\cite{jain2025bhasaanuvaad} developed by AI4Bharat\footnote{\url{https://ai4bharat.iitm.ac.in/}}, which contains approximately $44,\mathrm{k}$ hours of S$2$ST data with paired speech and text across $14$ Indic languages. From this collection, we select hi, ben, and urd, resulting in approximately $696$ hours of speech data. A summary of statistics for all training datasets is provided in Table~\ref{tab:dataset_stats} in Appendix~\ref{datatab}.

For evaluation, we employ both internal and external test sets. Internal evaluation sets are constructed by randomly sampling approximately $1\%$ of each training corpus, ensuring no overlap with training or validation data. For external benchmarking, we use the Few-shot Learning Evaluation of Universal Representations of Speech (FLEURS) dataset~\cite{conneau2023fleurs}, a widely adopted benchmark for speech translation, and select the zh-en translation subsets. Detailed statistics for all evaluation datasets are summarized in Table~\ref{tab:evaldatatable} in Appendix~\ref{datatab}.

\subsection{Baseline}
\label{baseline}
For system-level evaluation, we develop a cascaded speech translation baseline to benchmark the proposed DS$2$ST-LM model. The cascaded pipeline follows the conventional three-stage architecture comprising ASR, MT, and TTS. To keep the model computationally affordable, the DS$2$ST-LM framework incorporates the Whisper-small encoder, Qwen$2$--$0.5$B LLM, and CosyVoice vocoder. For a fair comparison, the same components are used in the cascaded system: Whisper-small~\cite{radford2023whisper} for ASR, Qwen$2$--$0.5$B~\cite{yang2024qwen2} for text translation, and CosyVoice~\cite{du2024cosyvoice} for speech synthesis.

All cascaded baseline modules (ASR, MT, and TTS) are fine-tuned using the same training corpus and hyperparameter configuration as DS$2$ST-LM. This ensures that observed performance differences arise solely from architectural differences between the cascaded and DS$2$ST-LM frameworks. 
We additionally construct an ST $+$ TTS baseline, in which Qwen-Audio~\cite{chu2024qwen2audio} directly generates target‑language text from source speech, followed by target-speech synthesis using the CosyVoice TTS system.

Beyond system-level baselines, we also establish data-level baselines using publicly available S$2$ST resources. Specifically, Seamless-Align~\cite{barrault2023seamlessm4t} is used to evaluate baseline performance under natural speech conditions. We further evaluate on the FLEURS dataset~\cite{conneau2023fleurs} to assess cross-domain generalization using a standardized benchmark.

\subsection{Training setup}
For model training, the Whisper-small encoder is employed as the audio feature extractor. The input speech is zero-padded to a fixed duration of $30$ seconds, encoded by the Whisper encoder, and subsequently projected to match the input dimensionality of the LLM. 
The Qwen$2$--$0.5$B model serves as the LLM backbone, while a group size of $G = 3$, as described in section~\ref{group-modelling}, is adopted to align the generation rates of text and semantic tokens. For speech synthesis, the CosyVoice vocoder is utilized with a codebook size of $4096$ to generate high-quality waveform outputs. During training, the Whisper encoder is kept frozen, whereas the LLM is fully fine-tuned to adapt to the speech translation task. The model is optimized using the Adam optimizer with a learning rate of $1\times10^{-4}$ and a batch size of $8$. A warm-up of $1000$ steps is applied, after which the learning rate is decayed according to a schedule with a decay factor $\gamma = 0.85$.  
Training is performed for a maximum $4$ epochs with an early stopping criterion based on validation loss saturation. Validation is conducted every $3000$ steps, and the best-performing checkpoint is selected based on the lowest validation and training loss. Mixed-precision (FP$16$) training is employed to improve memory efficiency and accelerate computation. During inference, greedy search decoding is used with a repetition penalty of $1.2$ to enhance generation diversity. All experiments are carried out on a system equipped with an Intel Xeon Gold $6338$N processor featuring $64$ CPU cores, $1$TB of system RAM, and an NVIDIA H$100$ GPU with $80$GB of GPU memory. 

\begin{table*}[!t]
\centering
\caption{Performance comparison of DS$2$ST-LM and baseline models using lexical and semantic evaluation metrics (BLEU, METEOR, BLEURT, COMET) across Chinese (zh)--English (en) datasets.}
\label{tab:comparison_metrics}
\renewcommand{\arraystretch}{1.4}
\setlength{\tabcolsep}{4.5pt}
\begin{tabular}{lcccccccccccc}
\hline
\multirow{2}{*}{\makecell{\textbf{Model /}\\\textbf{Datasets}}} &
\multicolumn{4}{c}{\textbf{Seamless-Align (zh--en)}} & 
\multicolumn{4}{c}{\textbf{GigaS$2$S-$1000$ (zh--en)}} & 
\multicolumn{4}{c}{\textbf{FLEURS (zh--en)}} \\
\cmidrule(lr){2-5} \cmidrule(lr){6-9} \cmidrule(lr){10-13}
& \textbf{BLEU} & \textbf{METEOR} & \textbf{BLEURT} & \textbf{COMET} 
& \textbf{BLEU} & \textbf{METEOR} & \textbf{BLEURT} & \textbf{COMET}
& \textbf{BLEU} & \textbf{METEOR} & \textbf{BLEURT} & \textbf{COMET} \\ 
\hline
Cascaded 
& $4.78$ & $0.25$ & $0.30$ & $0.34$
& $6.84$ & $0.16$ & $0.37$ & $0.39$
& $5.78$ & $0.23$ & $0.36$ & $0.38$ \\

ST $+$ TTS 
& $5.91$ & $0.27$ & $0.35$ & $0.49$
& $11.36$ & $0.32$ & $0.43$ & $0.54$
& $9.17$ & $0.25$ & $0.41$ & $0.53$ \\

DS$2$ST-LM 
& $7.11$ & $0.37$ & $0.42$ & $0.58$
& $\mathbf{14.71}$ & $\mathbf{0.45}$ & $\mathbf{0.53}$ & $\mathbf{0.71}$
& $11.46$ & $0.45$ & $0.53$ & $0.68$ \\
\hline
\end{tabular}
\end{table*}

\subsection{Evaluation Metric}
\label{evalmetric}






To evaluate the proposed DS$2$ST-LM framework, we employ a combination of objective and subjective metrics. Among these, BLEU~\cite{papineni2002bleu} evaluates n-gram overlap between system outputs and reference translations, reflecting lexical accuracy, while METEOR~\cite{banerjee2005meteor} incorporates synonym and stem matching to better capture semantic adequacy. We further incorporate semantic-oriented metrics, including BLEURT~\cite{sellam2020bleurt} and COMET~\cite{rei2020comet}, which are better aligned with human judgments and are robust to lexical variability commonly observed in LLM-generated translations. As these evaluation metrics are text-based, translated speech outputs are transcribed using the Whisper-large ASR model.

To evaluate speaker timbre preservation, we employ a WavLM-based~\cite{chen2022wavlm} speaker verification model to extract discriminative speaker embeddings. Speaker similarity is quantified using cosine similarity between embeddings of synthesized speech and reference speaker prompts, where higher values indicate stronger preservation of speaker identity. Perceptual speech quality is additionally assessed using Deep Noise Suppression Mean Opinion Score (DNSMOS)~\cite{reddy2021dnsmos}, a widely adopted speech quality estimator that predicts human Mean Opinion Score (MOS) ratings for speech naturalness and intelligibility.

In addition to objective evaluations, we conduct subjective listening tests based on the standard MOS~\cite{jia2019direct} to assess speech naturalness and speaker similarity. We also collect human judgments of adequacy and fluency~\cite{pal2023wmtindic} to evaluate semantic meaning preservation and grammatical well-formedness. A summary of all evaluation metrics is reported in Table~\ref{tab:eval_metrics} in Appendix~\ref{evalmetricA}.

\subsection{Results and Discussion}
In this subsection, we present and analyze the experimental results obtained from a comprehensive set of evaluations conducted on the proposed DS$2$ST-LM framework. 
\subsubsection{Baseline Performance Analysis}
As summarized in Table~\ref{tab:comparison_metrics}, we evaluate DS$2$ST-LM under three training configurations corresponding to different data conditions, as described in Section~\ref{baseline}. Across all datasets, DS$2$ST-LM is compared against the cascaded and ST $+$ TTS baselines using identical model components and training data. Unlike these cascaded approaches, DS$2$ST-LM jointly trains the speech encoder, projection module, LLM decoder, and vocoder in an end-to-end manner, enabling more coherent alignment between acoustic and semantic representations.

Because DS$2$ST-LM leverages a multilingual LLM, both lexical and semantic evaluation metrics are reported to provide a comprehensive assessment, as discussed in Section~\ref{evalmetric}. The results in Table~\ref{tab:comparison_metrics} show that the DS$2$ST-LM framework consistently outperforms the cascaded baselines across all metrics, including BLEU, METEOR, BLEURT, and COMET. These findings indicate that the end-to-end framework effectively mitigates the error propagation inherent in cascaded pipelines. Among all datasets, the best performance is achieved when DS$2$ST-LM is trained on the GigaS$2$ST-$1000$ corpus. This improvement is primarily attributed to the high quality of the dataset and its strong acoustic–semantic alignment. The corpus is built using high-accuracy MT, and the target speech is synthesized with a state-of-the-art TTS system in a single speaker voice. Consequently, the speech pairs are clean, consistent, and well aligned.

In contrast, the Seamless-Align corpus is curated from web-sourced audio, where bilingual parallel speech is aligned using automatic techniques such as forced alignment. As a result, the audio data exhibit substantial variability in recording conditions and inconsistencies in parallel alignment, negatively affecting system performance. These observations underscore the importance of carefully curated, acoustically parallel datasets for S$2$ST systems.

As LLM-based models may generate meaning-preserving translations using different lexical forms, semantic metrics offer a more reliable indicator of translation quality. A similar observation is evident in our results: although BLEU scores are slightly lower, DS$2$ST-LM achieves higher BLEURT and COMET scores. This difference highlights its ability to produce translations that are both semantically faithful and contextually accurate. 
\begin{table}[!t]
\centering
\caption{Impact of different projection architectures on the translation performance of DS$2$ST-LM trained on the GigaS$2$S-$1000$ dataset}
\label{tab:gigas2s_single}
\renewcommand{\arraystretch}{1.4}
\setlength{\tabcolsep}{4.5pt}
\begin{tabularx}{\columnwidth}{l *{4}{>{\centering\arraybackslash}X}}
\toprule
\textbf{Projector / Metrics} &
\textbf{BLEU} &
\textbf{METEOR} &
\textbf{BLEURT} &
\textbf{COMET} \\
\midrule
Linear                     
& $\mathbf{14.71}$ & $\mathbf{0.45}$ & $\mathbf{0.53}$ & $\mathbf{0.71}$ \\
Conv$1$D--Linear            
& $13.33$ & $0.44$ & $0.53$ & $0.70$ \\
Q-Former ($2$ layers)       
& $13.10$ & $0.44$ & $0.52$ & $0.70$ \\
Q-Former ($4$ layers)       
& $12.05$ & $0.43$ & $0.50$ & $0.67$ \\
\bottomrule
\end{tabularx}
\end{table}
\subsubsection{Effect of Projection Architectures}
We further analyze the influence of different projection architectures on model convergence behavior and translation quality by comparing three configurations: Linear, Conv$1$D–Linear, and Q-Former. The analysis focuses on learning stability, modeling capacity, and generalization performance. Two consistent trends emerge from the experiments. First, projection modules with higher representational capacity, such as Conv$1$D–Linear and Q-Former variants with two and four layers, exhibit noticeably faster convergence during training. Second, this rapid convergence is often followed by early saturation, leading to a decline in translation performance.

As summarized in Table~\ref{tab:gigas2s_single}, this behavior can be attributed to the fact that these expressive projectors introduce additional learnable temporal aggregation or semantic abstraction, which interferes with the fine-grained temporal alignment already encoded by the pretrained speech encoder. Unlike image processing, where spatial abstraction is often beneficial, speech representations require preserving frame-level temporal structure, as each frame carries fine-grained acoustic and linguistic information. Consequently, optimization with higher-capacity projectors tends to converge quickly to suboptimal solutions.
In contrast, the Linear projection module demonstrates more stable training behavior and achieves the highest translation scores across all evaluation metrics. By preserving temporal correspondence while providing sufficient dimensional alignment to the LLM, the Linear projector minimizes unnecessary transformations of encoder representations. These results indicate that, under the current training configuration and dataset scale, the Linear projection module offers the most robust balance between computational efficiency, convergence stability, and performance consistency.

Nevertheless, the Linear projection approach may exhibit limitations for certain tasks, as it does not explicitly model temporal context, suppress noise, or perform content selection. Moreover, its effectiveness depends on the quality of the pretrained speech encoder. In scenarios involving weak or noisy acoustic representations, very long speech inputs, or tasks that emphasize semantic reasoning, such as speech question answering or summarization, more expressive projectors like Conv$1$D–Linear or Q-Former may offer advantages.

\subsubsection{Semantic token: Speech vs Text}
There are two approaches to semantic token generation: speech-driven tokens produced by the $S^3$ tokenizer and text-driven tokens generated using a text-to-token LLM, as described in Section~\ref{token}. For the text-driven approach, the text-to-token LLM may generate different semantic token sequences for the same textual input depending on whether decoding is performed with or without cached attention states. To ensure token reproducibility, all LLM-based semantic tokens used in our experiments are generated without attention caching.

Experimental results in Table~\ref{tab:semantic_tokens} show that models conditioned on speech-derived semantic tokens outperform those using text-driven tokens when trained with the same data. Conditioning on speech-derived tokens generated from the GigaS$2$S-$1000$ dataset achieves a BLEU score of $14.71$, whereas replacing them with text-driven tokens results in a relative BLEU degradation of $17.28\%$. This observation is further validated on the Seamless-Align dataset under identical training conditions.

This discrepancy may be attributed to differences in the semantic embedding distributions induced by the  $S^3$ tokenizer and the text-to-token LLM. Although both tokenization strategies preserve equivalent semantic meaning, they differ in how that meaning is extracted. Speech-derived tokens capture how meaning is expressed in speech, including variations introduced by articulation and speaking style, whereas text-driven tokens represent a more standardized semantic form learned purely from text supervision.
\begin{table}[!t]
\centering
\caption{Effect of different semantic token generation techniques ($S^3$ vs.\ text-to-token LLM) on the translation performance of DS$2$ST-LM trained on the GigaS$2$S-$1000$ dataset.}
\label{tab:semantic_tokens}
\renewcommand{\arraystretch}{1.4}
\setlength{\tabcolsep}{4.5pt}
\begin{tabularx}{\columnwidth}{l *{4}{>{\centering\arraybackslash}X}}
\toprule
\textbf{Semantic token / Metrics} &
\textbf{BLEU} &
\textbf{METEOR} &
\textbf{BLEURT} &
\textbf{COMET} \\
\midrule
\textbf{Seamless-Align}   
&  &  &  &  \\
LLM--Toekns       
& $5.52$ & $0.31$ & $0.38$ & $0.43$ \\
S$3$--Tokens        
& $7.11$ & $0.37$ & $0.42$ & $0.58$ \\
\textbf{GigaS$2$S--$1000$}       
&  &  &  &  \\
LLM--Toekns       
& $12.58$ & $0.42$ & $0.50$ & $0.66$ \\
S$3$--Tokens        
& $\mathbf{14.71}$ & $\mathbf{0.45}$ & $\mathbf{0.53}$ & $\mathbf{0.71}$ \\
\bottomrule
\end{tabularx}
\end{table}

Despite the observed performance gap, text-driven semantic tokens can be a practical alternative when target speech data is unavailable. In such settings, widely available ST datasets can be leveraged to train direct S$2$ST models with only a modest degradation in performance. This approach substantially reduces the dependence on parallel S$2$ST corpora and enables the extension of direct S$2$ST models to language pairs where native S$2$ST data is scarce.

\subsubsection{Extensible S$2$ST system}
\begin{figure}[t]
\centering
\begin{tikzpicture}
\begin{axis}[
    width=\columnwidth,
    height=4.5cm,
    xlabel={Language Pairs},
    ylabel={COMET},
    xtick={1,2,3,4,5,6},
    xticklabels={
        \raisebox{0pt}[1.6ex][0.6ex]{fr--en},
        \raisebox{0pt}[1.6ex][0.6ex]{de--en},
        \raisebox{0pt}[1.6ex][0.6ex]{es--en},
        \raisebox{0pt}[1.6ex][0.6ex]{hi--en},
        \raisebox{0pt}[1.6ex][0.6ex]{ben--en},
        \raisebox{0pt}[1.6ex][0.6ex]{urd--en}
    },
    ymin=0.30,
    ymax=0.75,
    grid=major,
    legend style={
        at={(rel axis cs:0.40,0.70)},
        anchor=north east,
        font=\small,
        fill=none,
        draw=none
    }
]

\addplot+[color=blue, mark=o] coordinates {
    (1,0.43) (2,0.35) (3,0.45)
    (4,0.41) (5,0.34) (6,0.35)
};
\addlegendentry{Cascaded}

\addplot+[color=orange, mark=square*] coordinates {
    (1,0.63) (2,0.61) (3,0.65)
    (4,0.52) (5,0.45) (6,0.46)
};
\addlegendentry{ST $+$ TTS}

\addplot+[color=black, mark=triangle*] coordinates {
    (1,0.69) (2,0.65) (3,0.71)
    (4,0.71) (5,0.66) (6,0.54)
};
\addlegendentry{DS$2$ST-LM}

\end{axis}
\end{tikzpicture}
\caption{COMET score comparison of DS$2$ST-LM with cascaded models across six language pairs.}
\label{fig:comet_slope}
\end{figure}
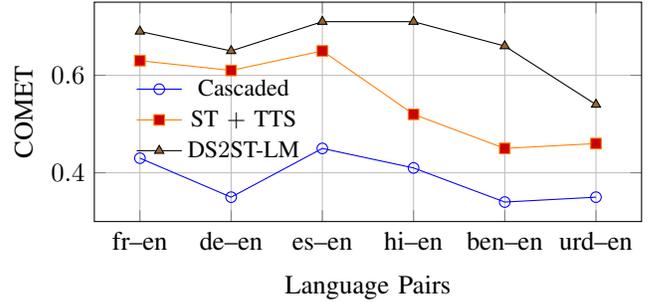

To assess the extensibility and generalization capability of the proposed DS$2$ST-LM framework, we extend the system to multiple language pairs, including Fr, De, and Es from the CVSS dataset, as well as Indic languages such as Hi, Ben, and Urd from the BHASAANUVAAD corpus. This multilingual evaluation facilitates a rigorous assessment of the model’s robustness across diverse linguistic typologies, phonetic structures, and acoustic conditions. As reflected by the BLEU and COMET scores reported in Figs.~\ref{fig:bleu_slope} and~\ref{fig:comet_slope}, DS$2$ST-LM consistently outperforms traditional cascaded and ST $+$ TTS baselines across all language pairs. Detailed results with additional metrics, including METEOR and BLEURT, are summarized in Appendix Tables~\ref{tab:covost2_metrics} and~\ref{tab:indic_bhasaanuvaad}.

The inferior performance of cascaded systems can be primarily attributed to error propagation across independently trained ASR, MT, and TTS modules. In particular, recognition errors at the ASR stage may misguide the downstream LLM-based translation process, thereby degrading the quality of synthesized speech. In contrast, DS$2$ST-LM jointly learns direct speech-to-speech mappings, thereby mitigating cumulative module-level errors and enabling tighter acoustic–semantic alignment between source and target languages.

Performance variations across languages can be explained by differences in the linguistic exposure of the underlying LLM during pre-training. Languages with greater representation in the pre-training corpus tend to achieve higher translation accuracy, whereas those with more limited exposure exhibit modest performance degradation. However, the consistent superiority of DS$2$ST-LM across both high and low-resource languages highlights its strong potential for scalable and multilingual speech translation.




\begin{figure}[t]
\centering
\begin{tikzpicture}
\begin{axis}[
    width=\columnwidth,
    height=4.5cm,
    xlabel={Language Pairs},
    ylabel={BLEU},
    xtick={1,2,3,4,5,6},
    xticklabels={
        \raisebox{0pt}[1.6ex][0.6ex]{fr--en},
        \raisebox{0pt}[1.6ex][0.6ex]{de--en},
        \raisebox{0pt}[1.6ex][0.6ex]{es--en},
        \raisebox{0pt}[1.6ex][0.6ex]{hi--en},
        \raisebox{0pt}[1.6ex][0.6ex]{ben--en},
        \raisebox{0pt}[1.6ex][0.6ex]{urd--en}
    },
    ymin=0,
    ymax=26,
    grid=major,
    legend style={
        at={(rel axis cs:0.40,0.62)},
        anchor=north east,
        font=\small,
        fill=none,
        draw=none
    }
]

\addplot+[color=blue, mark=o] coordinates {
    (1,3.97) (2,3.38) (3,4.19)
    (4,2.64) (5,1.68) (6,1.15)
};
\addlegendentry{Cascaded}

\addplot+[color=orange, mark=square*] coordinates {
    (1,19.69) (2,16.73) (3,20.18)
    (4,5.65) (5,3.56) (6,4.21)
};
\addlegendentry{ST $+$ TTS}

\addplot+[color=black, mark=triangle*] coordinates {
    (1,24.57) (2,19.09) (3,24.91)
    (4,22.82) (5,13.71) (6,5.41)
};
\addlegendentry{DS$2$ST-LM}

\end{axis}
\end{tikzpicture}
\caption{BLEU score comparison of DS$2$ST-LM with cascaded models across six language pairs.}
\label{fig:bleu_slope}
\end{figure}
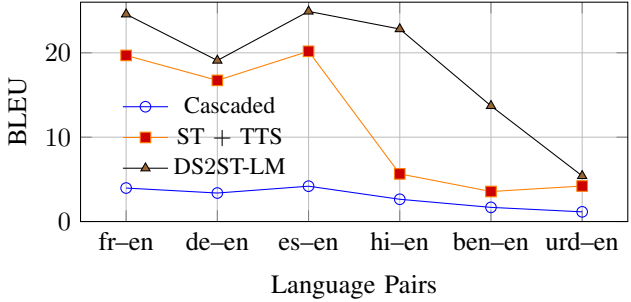

\subsubsection{Timbre controlable S$2$ST}
To assess the timbre preservation capability of the proposed DS$2$ST-LM system, we compare its speaker similarity performance with several direct S$2$ST models designed for speaker-identity retention. The evaluation is conducted on the fr-en test set, as speaker-similarity results for existing baselines are available for this language pair. In this setting, cosine similarity is computed between the speaker embedding of each translated utterance and the speaker embedding extracted from the corresponding reference prompt.

As reported in Table~\ref{tab:speaker_similarity}, DS$2$ST-LM achieves the highest speaker-similarity score (SIM) $(0.83)$, substantially outperforming TransVIP $(0.40)$, DA-Speech $(0.37)$, Translatotron$2$ $(0.43)$, and Translatotron $(0.32)$. These results demonstrate that integrating supervised semantic tokens with explicit speaker-embedding conditioning provides a robust mechanism for preserving speaker identity during translation. Unlike prior direct S$2$ST systems, where timbre drift often arises from the entangled linguistic and acoustic attributes in continuous latent spaces, DS$2$ST-LM conditions the vocoder on a fixed speaker embedding extracted from a short reference prompt. This explicit conditioning mitigates cross-lingual speaker drift and yields perceptually stable translated speech.

In addition to speaker similarity, we evaluate the perceptual quality of translated speech using the DNSMOS metric. As shown in Table~\ref{tab:speaker_similarity}, DS$2$ST-LM achieves a DNSMOS score of $3.54$, outperforming all comparison systems and approaching the naturalness level of the ground-truth audio $(3.86)$. Overall, these results indicate that DS$2$ST-LM effectively generates timbre-controlled speech, achieving superior performance in both speaker identity preservation and perceptual speech quality compared to existing direct S$2$ST systems.

\begin{table}[!t]
\centering
\caption{Comparison of speaker similarity and speech naturalness across different models to evaluate timbre preservation and perceptual speech quality on the fr--en evaluation set}
\label{tab:speaker_similarity}
\renewcommand{\arraystretch}{1.4}
\setlength{\tabcolsep}{4.5pt}
\begin{tabularx}{\columnwidth}{l >{\centering\arraybackslash}X >{\centering\arraybackslash}X}
\toprule
\textbf{Models / Speech Quality} &
\textbf{SIM} &
\textbf{DNSMOS} \\
\midrule
TransVIP          & $0.40$ & $3.26$ \\
DA-Speech         & $0.37$ & $3.32$ \\
Translatotron$2$  & $0.43$ & $3.41$ \\
Translatotron     & $0.32$ & $1.87$ \\
Ground Truth      & --     & $3.86$ \\
DS$2$ST-LM        & $\mathbf{0.83}$ & $\mathbf{3.54}$ \\
\bottomrule
\end{tabularx}
\end{table}

\section{Conclusion}
In this work, we introduce DS$2$ST-LM, a scalable single-stage direct S$2$ST framework that integrates a Whisper speech encoder, a learnable projection module, a Qwen$2$--$0.5$B LLM, and a timbre-controlled vocoder. To address the scarcity of parallel S$2$ST resources, we develop the GigaS$2$S-$1000$ corpus by extending GigaST with high-quality synthetic target speech generated using XTTS-v$2$. Through extensive experiments, we evaluated the behavior of the proposed system across different semantic token sources, projection designs, and language conditions. The results demonstrate that integrating a multilingual LLM with supervised semantic tokens and explicit timbre conditioning provides a stable and effective alternative to cascaded pipelines.

A few observations emerged from our analysis. First, semantic tokens derived directly from speech yield slightly higher translation accuracy than those generated using text-to-token LLMs. However, text-derived semantic tokens offer a practical alternative for direct S$2$ST by enabling training with widely available ST datasets. Second, although more expressive projectors, such as Conv$1$D–Linear and Q-Former, converge faster during training, the simple Linear projector exhibits the most stable behavior and achieves better generalization in the translation. The proposed system also demonstrated strong multilingual scalability, achieving competitive performance across seven language pairs, including those with limited S$2$ST resources. Additionally, incorporating a zero-shot timbre-control mechanism enables the model to synthesize target speech that preserves speaker identity, outperforming prior direct S$2$ST models in both speaker similarity and speech naturalness.

Despite these advances, the proposed approach remains constrained by the language coverage of the speech encoder, LLM, and vocoder, and it continues to face challenges when applied to low-resource languages. Future work will focus on scaling to alternative LLM architectures, jointly modeling acoustic and semantic tokens, and leveraging knowledge distillation to extend support to low-resource languages while reducing computational requirements. These directions can further enhance the scalability and practical utility of direct S$2$ST systems.




\bibliographystyle{IEEEtran}
\bibliography{mybib}
\clearpage
\appendix
\subsection{Subjective Evaluation}

Since the proposed system generates translated speech, subjective evaluation is conducted along two aspects: speech quality and translation quality. Speech quality is further evaluated in terms of speech naturalness and preservation of speaker characteristics, while translation quality is assessed based on meaning preservation and grammatical correctness. Accordingly, we adopt four subjective evaluation metrics: MOS Naturalness, MOS Speaker Similarity, Adequacy, and Fluency.

MOS Naturalness is used to assess the perceptual quality and naturalness of the synthesized translated speech, whereas MOS Speaker Similarity measures how closely the speaker characteristics of the translated speech resemble those of the reference speaker. Translation quality is evaluated using reference-based adequacy and fluency judgments. In this evaluation, fifteen linguistic experts evaluated $20$ randomly selected samples per model. All evaluations were conducted independently following predefined evaluation guidelines for MOS Naturalness, Speaker Similarity, Adequacy, and Fluency.
\subsubsection{MOS naturalness and speaker similarity}
To evaluate the speech synthesis quality of the translated output, we conduct a listening test using MOS Naturalness and MOS Speaker Similarity. For this evaluation, $20$ samples per model are randomly selected and listened to by the selected experts.

For MOS Naturalness, listeners rate the perceptual quality and naturalness of the synthesized translated speech on a five-point scale, where $1$ denotes bad, $2$ poor, $3$ fair, $4$ good, and $5$ excellent. For MOS Speaker Similarity, listeners compare the translated speech with the reference speaker prompt and assign a similarity score using the same five-point scale, reflecting how closely the speaker characteristics of the translated speech resemble those of the reference speaker.

The final MOS scores are obtained by averaging ratings across listeners and samples. As shown in Table~\ref{tab:MOS-nat-sim}, the proposed DS$2$ST-LM model is consistent with the objective evaluation results and outperforms the existing models in terms of speech naturalness and speaker similarity.

\begin{table}[!t]
\centering
\caption{MOS scores for speech naturalness (MOS Nat) and speaker similarity (MOS SIM) across different models on the Fr–En evaluation set}
\label{tab:MOS-nat-sim}
\renewcommand{\arraystretch}{1.4}
\setlength{\tabcolsep}{4.5pt}
\begin{tabularx}{\columnwidth}{l >{\centering\arraybackslash}X >{\centering\arraybackslash}X}
\toprule
\textbf{Models / Speech Quality} &
\textbf{MOS-SIM} &
\textbf{MOS-Nat} \\
\midrule
Translatotron     & $1.95$ & $3.05$ \\
Translatotron$2$  & $2.74$ & $3.37$ \\
DA-Speech         & $3.05$ & $3.29$ \\
TransVIP          & $3.70$ & $3.40$ \\
Ground Truth      & --     & $3.81$ \\
DS$2$ST-LM        & $3.95$ & $3.55$ \\
\bottomrule
\end{tabularx}
\end{table}
\subsubsection{Adequacy and Fluency}
We conduct a reference-based human evaluation to assess translation quality. Since all systems perform translation into English, the evaluation is carried out by annotators who are proficient in English. For each evaluation sample, annotators are provided with the reference English translation and the system-generated English transcription obtained from the translated speech using the Whisper-large ASR model.

Adequacy measures how well the system output preserves the semantic meaning of the reference translation. Annotators compare the system output with the reference text and assign an adequacy score on a five-point scale, where $1$ denotes incorrect meaning, $2$ little meaning preserved, $3$ partially missing meaning, $4$ mostly preserved meaning, and $5$ complete meaning preservation.

Each annotator evaluates $20$ samples per model, and the finFluency is evaluated using the same five-point scale and measures the grammatical correctness and naturalness of the generated English text. The results are reported in Table~\ref{tab:Adequacy_Fluency}, where the trends observed in human evaluation are consistent with those of the objective evaluation metrics.

\begin{table}[!t]
\centering
\caption{Adequacy and Fluency scores across models for evaluating semantic and grammatical correctness of translated speech}
\label{tab:Adequacy_Fluency}
\renewcommand{\arraystretch}{1.4}
\setlength{\tabcolsep}{4.5pt}
\begin{tabularx}{\columnwidth}{l >{\centering\arraybackslash}X >{\centering\arraybackslash}X}
\toprule
\textbf{Models / Score} &
\textbf{Adequacy} &
\textbf{Fluency} \\
\midrule
zh-en         & $3.26$ & $3.95$ \\
fr–en         & $3.38$ & $4.01$ \\
de–en         & $2.95$ & $3.54$ \\
es–en         & $3.34$ & $3.96$ \\
hi–en         & $3.12$ & $3.78$ \\
ben–en        & $2.46$ & $3.18$ \\
urd–en        & $1.62$ & $2.88$ \\
\bottomrule
\end{tabularx}
\end{table}

\subsection{Evaluation Metric}
\label{evalmetricA}
A brief summary of the evaluation metrics used in this study is provided in Table~\ref{tab:eval_metrics}.
\begin{table}[ht]
\centering
\caption{Evaluation metrics for assessing translation quality (text) and perceptual speech quality (audio)}
\label{tab:eval_metrics}

\setlength{\tabcolsep}{4.5pt}
\renewcommand{\arraystretch}{1.4}

\begin{tabularx}{\columnwidth}{%
  >{\arraybackslash}p{1.9cm}   
  >{\arraybackslash}X          
}
\hline
\textbf{Metric} & \textbf{Description / Target} \\ 
\hline

\textbf{BLEU~\cite{papineni2002bleu}} & n-gram overlap precision. \\
\textbf{METEOR~\cite{banerjee2005meteor}} & BLEU with synonym, stem, and word-order matching. \\
\textbf{BLEURT~\cite{sellam2020bleurt}} & Contextual semantic similarity (BERT-based). \\
\textbf{COMET~\cite{rei2020comet}} & Neural semantic similarity (LM-based). \\
\textbf{Speaker Similarity~\cite{chen2022wavlm}} & Cosine similarity between speaker embeddings. \\
\textbf{DNMOS~\cite{reddy2021dnsmos}} & Neural naturalness score ($1-5$ scale). \\ 
\textbf{MOS\cite{jia2019direct}} & Subjective naturalness score ($1-5$ scale). \\ 
\textbf{Adequacy~\cite{pal2023wmtindic}} & Human judgment of meaning preservation ($1-5$). \\
\textbf{Fluency~\cite{pal2023wmtindic}} & Human judgment of grammatical naturalness ($1-5$). \\ \hline

\end{tabularx}

\end{table}

\subsection{Dataset Details}
Statistics of the training datasets used in this study are reported in Table~\ref{tab:dataset_stats}, and detailed information about the evaluation sets is presented in Table~\ref{tab:evaldatatable}.
\label{datatab}
\begin{table}[!t]
\centering
 \caption{Statistics summarizing the number of sentences (in thousands, K) and total speech duration (in hours, h) across languages (Lang) and datasets (Data).}
\label{tab:dataset_stats}

\setlength{\tabcolsep}{4.5pt}
\renewcommand{\arraystretch}{1.4}

\begin{tabularx}{\columnwidth}{
  >{\centering\arraybackslash}p{0.8cm}
  >{\centering\arraybackslash}X
  >{\centering\arraybackslash}X
  >{\centering\arraybackslash}X
  >{\centering\arraybackslash}X
}
\hline
\textbf{Lang / Data} &
\textbf{GigaS$2$ST-$1000$} &
\textbf{Seamless-Align} &
\textbf{CVSS} &
\textbf{Bhasaanubad} \\
\cline{1-5}
zh & 
\mbox{$833$K \textbar{} $1000$h} & 
\mbox{$219$K \textbar{} $1000$h} & 
-- & -- \\

fr   & -- & -- & 
\mbox{$145$K \textbar{} $180$h} & -- \\

es  & -- & -- & 
\mbox{$77$K \textbar{} $86$h} & -- \\

de   & -- & -- & 
\mbox{$85$K \textbar{} $118$h} & -- \\

hi    & -- & -- & -- & 
\mbox{$226$K \textbar{} $449$h} \\

ben  & -- & -- & -- & 
\mbox{$78$K \textbar{} $172$h} \\

urd     & -- & -- & -- & 
\mbox{$23$K \textbar{} $75$h} \\
\hline
\end{tabularx}

\end{table}

\begin{table}[!t]
\centering
\caption{Statistics of the evaluation set summarizing the number of sentences and total speech duration (in hours, h) across languages (Lang) and datasets (Data).}
\label{tab:evaldatatable}

\setlength{\tabcolsep}{4.5pt}
\renewcommand{\arraystretch}{1.4}

\begin{tabularx}{\columnwidth}{
  >{\centering\arraybackslash}p{0.8cm}
  >{\centering\arraybackslash}X
  >{\centering\arraybackslash}X
  >{\centering\arraybackslash}X
  >{\centering\arraybackslash}X
  >{\centering\arraybackslash}X
}
\hline
\textbf{Lang / Data} &
\textbf{GigaS$2$ST-$1000$} &
\textbf{Seamless-Align} &
\textbf{CVSS} &
\textbf{Bhasaanubad} &
\textbf{Fleurs} \\
\cline{1-6}

zh & 
\mbox{$8330$ \textbar{} $21$h} & 
\mbox{$2194$ \textbar{} $6$h} & 
-- & -- & 
\mbox{$349$ \textbar{} $1.1$h} \\

fr   & -- & -- & 
\mbox{$1453$ \textbar{} $2.3$h} & -- & -- \\

es  & -- & -- & 
\mbox{$755$ \textbar{} $1.6$h} & -- & -- \\

de   & -- & -- & 
\mbox{$853$ \textbar{} $1.7$h} & -- & -- \\

hi    & -- & -- & -- & 
\mbox{$2267$ \textbar{} $4.8$h} & -- \\

ben  & -- & -- & -- & 
\mbox{$751$ \textbar{} $1.6$h} & -- \\

urd     & -- & -- & -- & 
\mbox{$230$ \textbar{} $0.7$h} & -- \\

\hline
\end{tabularx}
\end{table}

\subsection{Results for Extensible S2ST system}
We extend DS$2$ST-LM to multiple language pairs, including fr, de, es, hi, ben, and urd. Detailed results are reported in Tables~\ref{tab:covost2_metrics} and~\ref{tab:indic_bhasaanuvaad}.
\begin{table*}[!b]
\centering
\caption{Performance of DS$2$ST-LM across three language pairs (fr--en, de--en, and es--en) from the CVSS dataset, evaluated using lexical and semantic metrics (BLEU, METEOR, BLEURT, COMET).}
\label{tab:covost2_metrics}
\renewcommand{\arraystretch}{1.4}
\setlength{\tabcolsep}{4.5pt}
\begin{tabularx}{\textwidth}{l *{12}{>{\centering\arraybackslash}X}}
\hline
\multirow{2}{*}{\makecell{\textbf{Model /}\\\textbf{Languages}}} &
\multicolumn{4}{c}{\textbf{fr--en}} &
\multicolumn{4}{c}{\textbf{de--en}} &
\multicolumn{4}{c}{\textbf{es--en}} \\
\cmidrule(lr){2-5} \cmidrule(lr){6-9} \cmidrule(lr){10-13}
& \textbf{BLEU} & \textbf{METEOR} & \textbf{BLEURT} & \textbf{COMET}
& \textbf{BLEU} & \textbf{METEOR} & \textbf{BLEURT} & \textbf{COMET}
& \textbf{BLEU} & \textbf{METEOR} & \textbf{BLEURT} & \textbf{COMET} \\ \hline
Cascaded  
& $3.97$ & $0.25$ & $0.37$ & $0.43$ 
& $3.38$ & $0.21$ & $0.29$ & $0.35$ 
& $4.19$ & $0.27$ & $0.41$ & $0.45$ \\
ST $+$ TTS  
& $19.69$ & $0.49$ & $0.55$ & $0.63$ 
& $16.73$ & $0.40$ & $0.50$ & $0.61$ 
& $20.18$ & $0.48$ & $0.52$ & $0.65$ \\
DS$2$ST-LM  
& $\mathbf{24.57}$ & $\mathbf{0.50}$ & $\mathbf{0.56}$ & $\mathbf{0.69}$ 
& $\mathbf{19.09}$ & $\mathbf{0.41}$ & $\mathbf{0.51}$ & $\mathbf{0.65}$ 
& $\mathbf{24.91}$ & $\mathbf{0.49}$ & $\mathbf{0.54}$ & $\mathbf{0.71}$ \\ \hline
\end{tabularx}
\end{table*}

\begin{table*}[!b]
\centering
\caption{Performance of DS$2$ST-LM across three language pairs (hi--en, ben--en, and urd--en) from the BHASAANUVAAD dataset, evaluated using lexical and semantic metrics (BLEU, METEOR, BLEURT, COMET).}
\label{tab:indic_bhasaanuvaad}
\renewcommand{\arraystretch}{1.4}
\setlength{\tabcolsep}{4.5pt}
\begin{tabularx}{\textwidth}{l *{12}{>{\centering\arraybackslash}X}}
\toprule
\multirow{2}{*}{\makecell{\textbf{Model /}\\\textbf{Languages}}} &
\multicolumn{4}{c}{\textbf{hi--en}} &
\multicolumn{4}{c}{\textbf{ben--en}} &
\multicolumn{4}{c}{\textbf{urd--en}} \\
\cmidrule(lr){2-5} \cmidrule(lr){6-9} \cmidrule(lr){10-13}
& \textbf{BLEU} & \textbf{METEOR} & \textbf{BLEURT} & \textbf{COMET}
& \textbf{BLEU} & \textbf{METEOR} & \textbf{BLEURT} & \textbf{COMET}
& \textbf{BLEU} & \textbf{METEOR} & \textbf{BLEURT} & \textbf{COMET} \\ 
\midrule
Cascaded  
& $2.64$ & $0.18$ & $0.32$ & $0.41$ 
& $1.68$ & $0.16$ & $0.31$ & $0.34$ 
& $1.15$ & $0.16$ & $0.24$ & $0.35$ \\
ST $+$ TTS 
& $5.65$ & $0.23$ & $0.33$ & $0.52$ 
& $3.56$ & $0.19$ & $0.28$ & $0.45$ 
& $4.21$ & $0.22$ & $0.30$ & $0.46$ \\
DS$2$ST-LM  
& $\mathbf{22.82}$ & $\mathbf{0.45}$ & $\mathbf{0.49}$ & $\mathbf{0.71}$ 
& $\mathbf{13.71}$ & $\mathbf{0.34}$ & $\mathbf{0.41}$ & $\mathbf{0.66}$ 
& $\mathbf{5.41}$  & $\mathbf{0.31}$ & $\mathbf{0.34}$ & $\mathbf{0.54}$ \\
\midrule
\end{tabularx}
\end{table*}
\vspace{\baselineskip}



 




\vfill

\end{document}